# Electric Field Control of Magnetism of Mn dimer supported on Carbon-doped-h-BN surface

Mihir Ranjan Sahoo[1,2], Saroj Kumar Nayak[2], and Kalpataru Pradhan[3,*]

[1]Harish-Chandra Research Institute, Chhatnag Road, Jhunsi, Allahabad, India-211019

[2]School of Basic Sciences, Indian Institute of Technology Bhubaneswar, India -752050

[3]Theory Division, Saha Institute of Nuclear Physics, HBNI, Kolkata, India-700064

[*] Email- kalpataru.prdhan@saha.ac.in

Abstract:

Using density functional theory we show that the interaction between two Mn atoms can be tuned from anti-ferromagnetic (AFM) to ferromagnetic (FM) state by creating charge disproportion between the two on a 2D surface. The non-metallic planar heterostructures, the 2D surface, in our work is designed by doping carbon hexagon rings in a hexagonal boron nitride (h-BN) sheet. In addition, we show that an external electric field can be used to control the charge disproportion and hence the magnetism. In fact, our calculations demonstrate that the magnetic states of the dimer can be switched from AFM to FM or vice versa in an external electric field. The origin of this magnetic switching is explained using the charge transfer from (or to) the Mn dimer to (or from) the 2D material. The switching between anti-ferromagnetic to ferromagnetic states can be useful for future spintronic applications.

Introduction:

The significant and distinct electronic, magnetic, and transport properties exhibited in low-dimensional systems due to electronic confinement have gained huge technological interest for designing smaller and smarter electronic devices[1–4]. In the era of global digitalization, the necessity of storing data and information generated due to the enormous use of high-performance computing, multipurpose advanced cellular devices, have stimulated tremendous scientific efforts to engineer advanced nanoscale-spintronics devices. For this, molecular magnets, that can be operated at high speed with low power consumption, are considered as potential information-storage units[5–9]. Apart from the technological point of view, understanding the behavior of macroscopic magnetism at reduced dimensionality and manipulating it through external means is also interesting for basic scientific explorations[10–12].

In the spin or memory devices, reversing magnetization through powerful, fast and tunable techniques is very important in writing process[13,14]. This control and tuning of magnetization of the system can be achieved by switching one magnetic state to another through various external sources like magnetic field[15], laser field[16,17], temperature[18], pressure[15,19–21], spin-polarized current[22–24]. However, these approaches for magnetism manipulations are not efficient as they act non-locally which can affect the neighboring units. Thus, the search for alternative and efficient technique through which magnetism can be controlled and manipulated at nanoscale brought significant attentions towards the application of external electric field (EEF). From experimental point of view, through the tip of scanning tunneling microscope the EEF can be applied locally to the system. Previous experimental studies reported that magnetization, magnetic exchange interactions, and magnetic anisotropy can be controlled and modified by EEF in multiferroic heterostructures[25–27] (due to coupling between electric field with magnetization through electric polarization), metal thin films[28–34] (due to shifting of the Fermi level at the interfaces), semicondcutors[35–37] (due to change in charge carrier density), nanomagnets[38], magnetic tunnel junctions(MTJs)[39,40] (due to change in charge carrier population[41]), magnetic nanomesh[42]. In addition, first principles calculations also reported that the magnetism exhibited in the nanostructures like magnetic clusters, MTJs, metallic surfaces, metallic thin films etc. can be controlled through EEF[6,39,43–47]. However, manipulation of magnetism in TM cluster above a 2D non-metallic planar heterostructures by simultaneous variation of EEF and concentration fraction of the substrate has rarely been reported.

Among the various magnetic storage devices, designed with low-dimensionality, the magnetic nano-clusters supported on different non-magnetic two-dimensional (2D) surfaces are convenient for preparing portable and accessible devices with high storage density[9,48]. It is important to mention here that *3d* transition metal (TM) clusters deposited over heavy TM surfaces have shown high magnetic anisotropy energy (MAE) due to presence of strong spin-orbit coupling and induction of spin-polarization created from the hybridization of *3d* electrons of TM with *4d* or *5d* elements[48–50]. However, heavy electron-electron scattering makes the spin life span of these devices very short which hinders the stability of the magnetization[14]. In this regards, graphene[51,52], the mostly used 2D crystalline structure, is considered as a promising spintronics material due to its high charge-carrier mobility, long spin diffusion length due to very low spin-orbit coupling, observation of quantum Hall effect at room temperature and unique Dirac cone band structures[53–56]. On the other hand, the search for an alternate 2D crystalline substrate for magnetic nano-clusters also focuses on hexagonal boron nitride (h-BN) monolayers (insulating counterpart of the graphene)[57,58] which shows high temperature resistance. In addition, for designing of 2D metal free substrate with high thermal stability, doping of light element such as carbon atoms in h-BN considered to be an efficient method[59–62]. The electronic, magnetic and catalytic properties of these heterostructures formed by substitution of carbon atoms in h-BN domain can be modified by geometry, concentration and configuration of atomic-level doping and make them key candidates for potential applications[63–66]. From both first principles and experimental studies, it is reported that carbon hexagon rings (graphene quantum dots) with different sizes and distributions inside h-BN monolayer reduces the band gap, which adds another functionality in to the system[67–71]. In this context, magnetic cluster placed above the planar heterostructure of graphene quantum dots doped in h-BN (C-doped-h-BN) is not only interesting from fundamental research but also provides new perspectives for specific applications. In this work, we have considered manganese dimer placed above the C-doped-hBN surface to investigate the magnetic states of the dimer with respect to the concentration of graphene rings in h-BN sheet. In addition we show that an external electric field can be used to control the magnetism, which can be used for future spintronics applications.

The magnetic states of $Mn_2$ and $Mn_2^+$ are quite interesting where transfer of electrons from Mn can switch the magnetic states. Using Hartee-Fock calculation and Heisenberg exchange interaction in 1964, Nesbet found that the antiferromagnetic coupling of $Mn_2$ molecule is energetically stable than the ferromagnetic state[72]. From early experiments, it was also reported that $Mn_2$ molecules placed in cyclopropane or argon matrices exhibits antiferromagnetic exchange coupling[73,74]. On the other hand, an experiment performed by Van Zee et al. showed that $Mn_2^+$ ion was found to be ferromagnetic with a shorter bond length[75]. Keeping all these facts in mind we have placed $Mn_2$ dimer on C-doped-h-BN monolayers to investigate the coupling between the Mn atoms. The ground state is ferromagnetic or antiferromagnetic is very much dependent upon the charge transfer from $Mn_2$ to C-doped-h-BN surface. Interestingly charge transfer depends upon the band gap of the C-doped-h-BN surface. Then we report that the application of EEF can switch the magnetization in the Mn dimer placed above C-doped-h-BN layer. So the objective of our investigation is primarily two-fold: (1) To show that the switching between two magnetic states in the dimer can be achieved through variation of carbon concentration in h-BN substrate and (2) To explore the nature of magnetic coupling and charge transfer in presence of external electric field.

Numerical Methods:

The first principles calculations based on density functional theory were performed to study the structural, electronic, and magnetic properties of Mn dimer on h-BN and C-doped-h-BN surfaces. All the results shown in this work obtained by using the projector augmented wave (PAW)[76] method implemented in the Vienna *ab initio* Simulation Package (VASP) code[77,78]. The contribution of exchange-correlation functional towards total energy functional was considered by using PW91 functional[79]. Due to the presence of Mn atoms, we included Hubbard type on-site Coulomb potential U with generalized gradient approximations (GGA+U) in the calculations. By using the formulation proposed by Dudarev & Botton[80], we have considered U= 5.0 eV due to the presence of localized d-orbitals Mn atom. The plane-wave basis set was expanded with kinetic energy cutoff of 500 eV.

Through the conjugate gradient method, atoms of all the structures were relaxed until the forces on atoms were less than 0.001 eV/Å. For cluster calculations, we put Mn dimers and $Mn_2X$ clusters (X= H, Cl, F) in a box of dimensions 20Å×20Å×20Å. The spin polarization calculations were performed for Mn dimer placed above 8×8 supercell of pristine h-BN and C-doped-h-BN monolayer (XY plane) and a vacuum of 20Å was maintained in z-direction to avoid the interaction between images created due to periodicity. The Brillouin zone was sampled in Gamma-centered method with k-mesh grid of 2×2×1 for the system containing pristine h-BN/C-doped-h-BN monolayer (8×8×1 supercell) self-consistent calculations whereas 1×1×1 k-mesh grid is enough for calculations of clusters. A perpendicular dipole sheet along XY-plane is introduced at the center of the supercell to simulate external electric field along z-direction. A series of external electric fields (EEF) are applied along both the vertical directions (positive and negative z-axis) to study its effect on the charge transfer and the magnetism. To avoid the artificial long-range Coulomb interactions due to presence of EEF, a dipole correction was taken into account in the calculations. Moreover, the distribution and transfer of charges between the atoms in different systems were estimated with the help of Bader charge analysis[81,82].

Results and Discussions:

We started our calculations with Mn dimer where the ground state is expected to be AFM and then gradually move to analyze the ground state magnetic properties of different triatomic clusters ($Mn_2Cl$, $Mn_2H$, and $Mn_2F$). The ground state geometries (with magnetic moments and bond lengths) corresponding to FM and AFM states for $Mn_2$, $Mn_2Cl$, $Mn_2H$, $Mn_2F$ are given in Fig.1. The Mn dimer prefers the AFM state, which is 28 meV lower than that of the FM state. We define this energy difference between the AFM and the FM states as the exchange energy. The bond length in AFM state (3.31Å) is slightly smaller than the bond length in FM state (3.35Å). In this case, the total magnetic moment of the dimer is equal to 10μB and 0μB in FM and AFM states respectively. Then we switch to analyze the ground state magnetic moment of different triatomic clusters ($Mn_2Cl$, $Mn_2H$ and $Mn_2F$). When one Cl atom is attached to Mn dimer, then the cluster is stable in the FM configuration with energy 300 meV lower than the AFM configuration. The total magnetic moments of $Mn_2Cl$ cluster in FM and AFM states are 11 μB and 1 μB respectively. So the addition of Cl atom switches the AFM configuration of $Mn_2$ dimer to FM configuration, similar to earlier calculations[83]. With the help of Bader charge analysis, we calculated the amount of charge transfer at each atom of the clusters and found that the charge of amount 0.73e is transferred from Mn atoms to Cl atoms. Due to electronegativity of Cl atom, it draws electron from Mn and makes the Mn dimer similar to $Mn_2^+$ that favors in FM states with magnetic moment of 11μB. The bond length between Mn atoms decreases to 2.91 Å. Note that the bond length of $Mn_2^+$ is shorter than $Mn_2$ as mentioned earlier. Recent theoretical and experimental works show that the observed FM state in di-nuclear TM complexes can be explained by double-exchange interactions that arise due to the charge transfer and/or charge disproportionation between the TM atoms[84–87]. We will discuss more about the competition of double exchange and super exchange interactions between Mn atoms in $Mn_2Cl$ later.

In order study the effect of the charge transfer on the magnetization and the exchange energy, we further studied by adding atoms having different electronegativity to the Mn dimer. In $Mn_2H$, though H is less electronegative than Cl, there is still considerable charge (0.53e) transfer from Mn atoms to H atom and as a result Mn dimer is found to be FM with magnetic moment 11 μB. But the exchange energy decreases to 180 meV in $Mn_2H$. Then we investigated with a higher electronegative atom i.e., F to see the charge transfer and ground state configuration. The amount of charge transfer from Mn dimer towards F in $Mn_2F$ is higher (0.79e) than above two scenarios. As expected the $Mn_2F$ cluster attains FM ground state with larger exchange energy (360 meV). This shows that the amount of charge transfer controls the exchange energy.

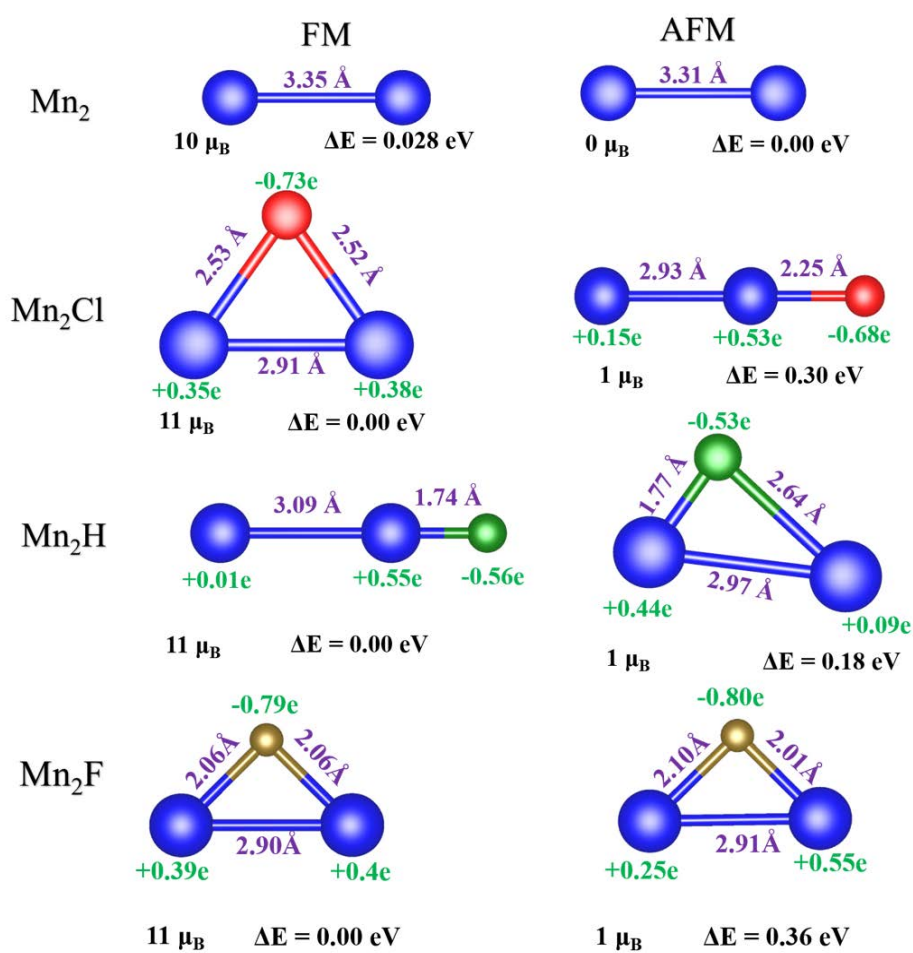

**Figure 1** Equilibrium geometrical structures of $Mn_2$ and $Mn_2X$ (X = Cl, H, F) clusters in both ferromagnetic (FM) and antiferromagnetic (AFM) states with bond lengths. ΔE represents the energy difference from the corresponding ground state energy. Magnetic moment of each structure is given in Bohr's magneton (μB). Blue spheres represent Mn atoms.

From the above results, it is clear that the charge transfer plays an integral role in deciding the magnetic states of the Mn dimer. With this information, we have designed a model analogous to triatomic cluster by introducing C-doped-h-BN 2D surface instead of electronegative atoms. Mn dimer is placed above C-doped-h-BN surface to find the relationship between the charge transfer and the exchange energy. A high potential barrier due to large band gap in h-BN monolayer[88,89] may inhibit the charge transfer from Mn dimer to the h-BN and as a result the AFM state is expected. In fact we find an AFM ground state, which will be discussed later. Can we tune the band gap by doping h-BN to control the charge transfer and hence the magnetism? In order to test this idea, first we focused to reduce the band gap of pristine h-BN monolayer by doping carbon hexagon rings of various sizes as shown in the Fig. 2. Here, we have calculated the band structures of 4, 9, and 16 hexagonal rings of carbon clusters doped inside 8x8 h-BN supercell and compared the results with pristine h-BN band structures. For convenience we named these in-plane 2D heterostructures as 4C-ring-h-BN, 9C-ring-h-BN, and 16C-ring-h-BN surfaces respectively in this paper and the structures are shown in Fig. 3. When we increase the amount of carbon hexagon rings in h-BN, the electronic band gap gradually decreases (Fig. 2). More the number of carbon rings embedded in h-BN sheet, more the states occupied in the gap of h-BN and hence smaller the bandgap of the composite system[71].

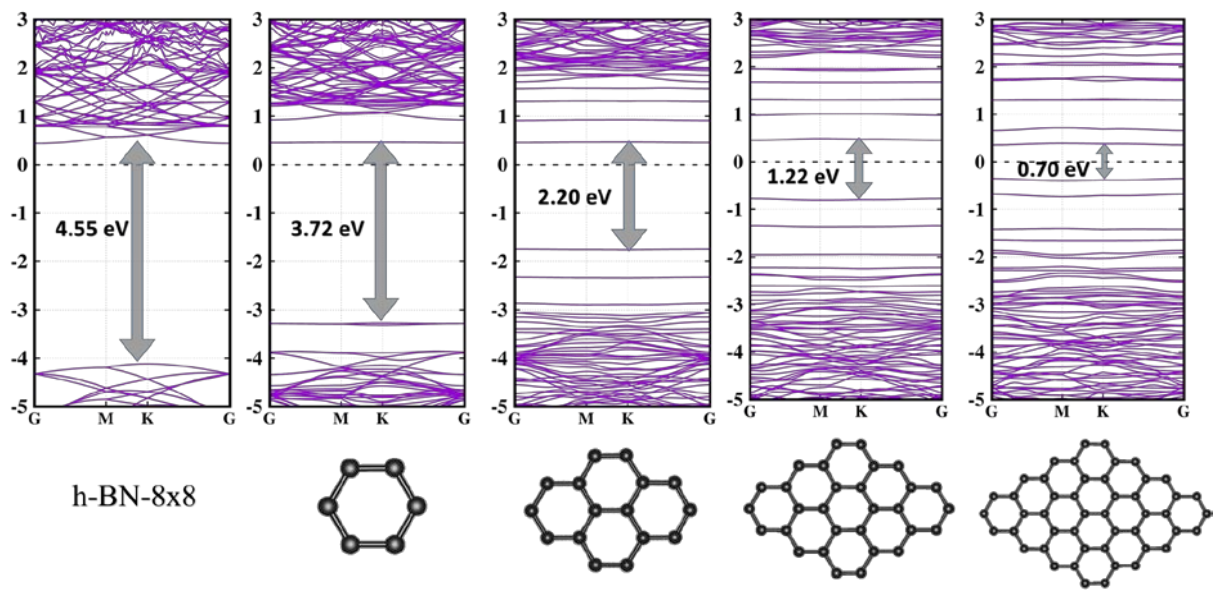

**Figure 2** Electronic band structures of h-BN and C-doped-h-BN structures. Size and shape of carbon hexagon rings embedded in h-BN are also shown. Please see Fig. 3 for a more elaborated picture.

Next, we have studied magnetic properties of $Mn_2$ on pristine h-BN and C-doped-h-BN surfaces. For this we calculated the stable geometrical orientation of Mn dimer above pristine h-BN and C-doped-h-BN among various positions. There are different possible geometries in which Mn dimer can be placed above the honeycomb lattice and lowest four isomers are shown in the Fig. 3. Among the four configurations (S1, S2, S3, and S4), we found the S3 configurations are the energetically stable structure for all surfaces. In presence of h-BN monolayer, Mn dimer remains AFM in nature with exchange energy 25.1 meV. The charge transfer from Mn dimer to the h-BN surface is minimal (0.06e) while Mn-Mn bond length is found to be 3.29 Å. It is important to note that bond length between Mn atoms in Mn$_2$ molecule is 3.31 Å (see Fig. 1). Similarly, AFM is the ground state for Mn dimer on 4C-ring-h-BN and bond length is found to be 3.35Å. However, the system containing Mn dimer above 9C-ring-h-BN is FM in nature though the exchange energy is very less (13 meV) and the bond length between the Mn atoms decreases to 3.30 Å [Fig. 4(a)]. However, above the 16C-ring-h-BN substrates, Mn dimer prefers to be in FM state with considerably lower energy (134 meV) than the corresponding AFM state with comparatively shorter bond length (3.12Å). So the distance between Mn atoms is minimum on 16C-ring-h-BN and the average vertical distance between Mn atoms and the 2D surface is also small. The average distance between Mn atoms and the 2D surface in case of 16C-ring-h-BN is 2.94Å while it is 3.41Å for h-BN case. It is clear from Fig. 4(b) that by increasing the C-rings, the value of exchange energy ($E_{FM} - E_{AFM}$) is gradually increasing and the system is tending towards FM state albeit in absence any external electric field. To correlate the FM states of Mn dimer on 9C-ring-h-BN and 16-C-ring-BN surfaces with the amount of charge transfer, we have performed the Bader charge analysis for each structure. From Fig. 4(c), it is seen that the charge transfer from the Mn dimer towards pristine h-BN monolayer is the least i.e. 0.06e whereas it is highest (0.34e) in the presence of 16C-ring-h-BN. Therefore, all these results indicate that by tuning the band gap of the substrate, we can enhance the charge transfer and as result an AFM system switches to a FM system.

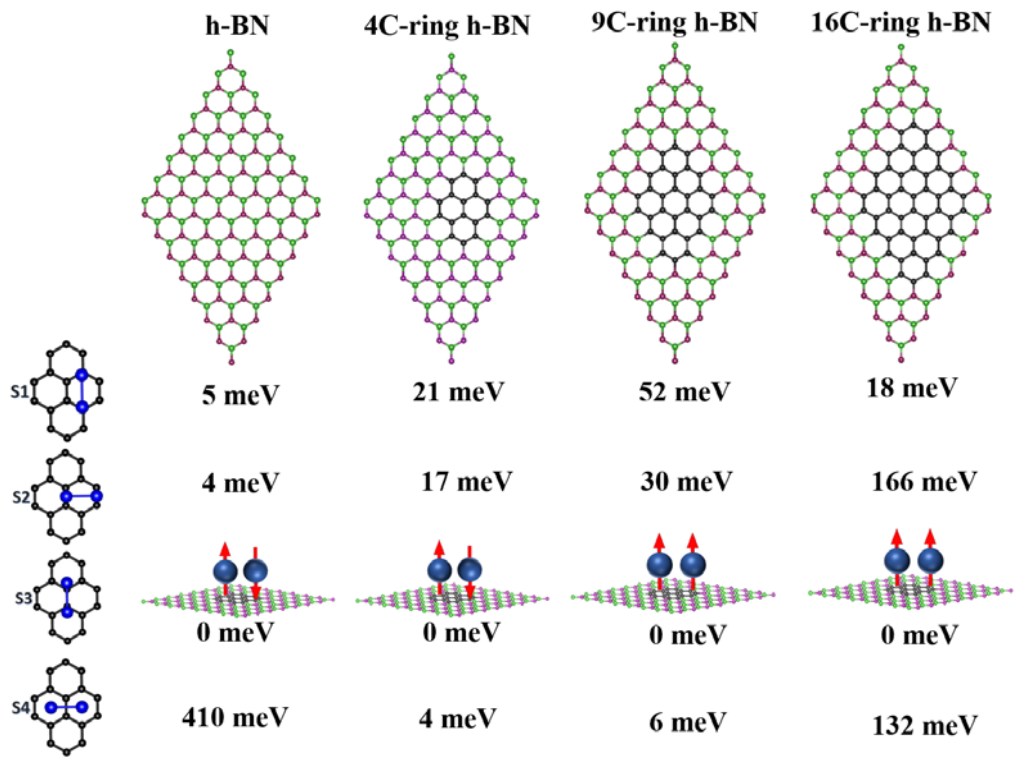

**Figure 3** Relative energy difference of various orientation of Mn$_2$ dimer placed on h-BN and C-doped h-BN substrate with respect to stable configurations. Green, magenta, black, and blue spheres represent boron, Nitrogen, Carbon, and Manganese atoms respectively.

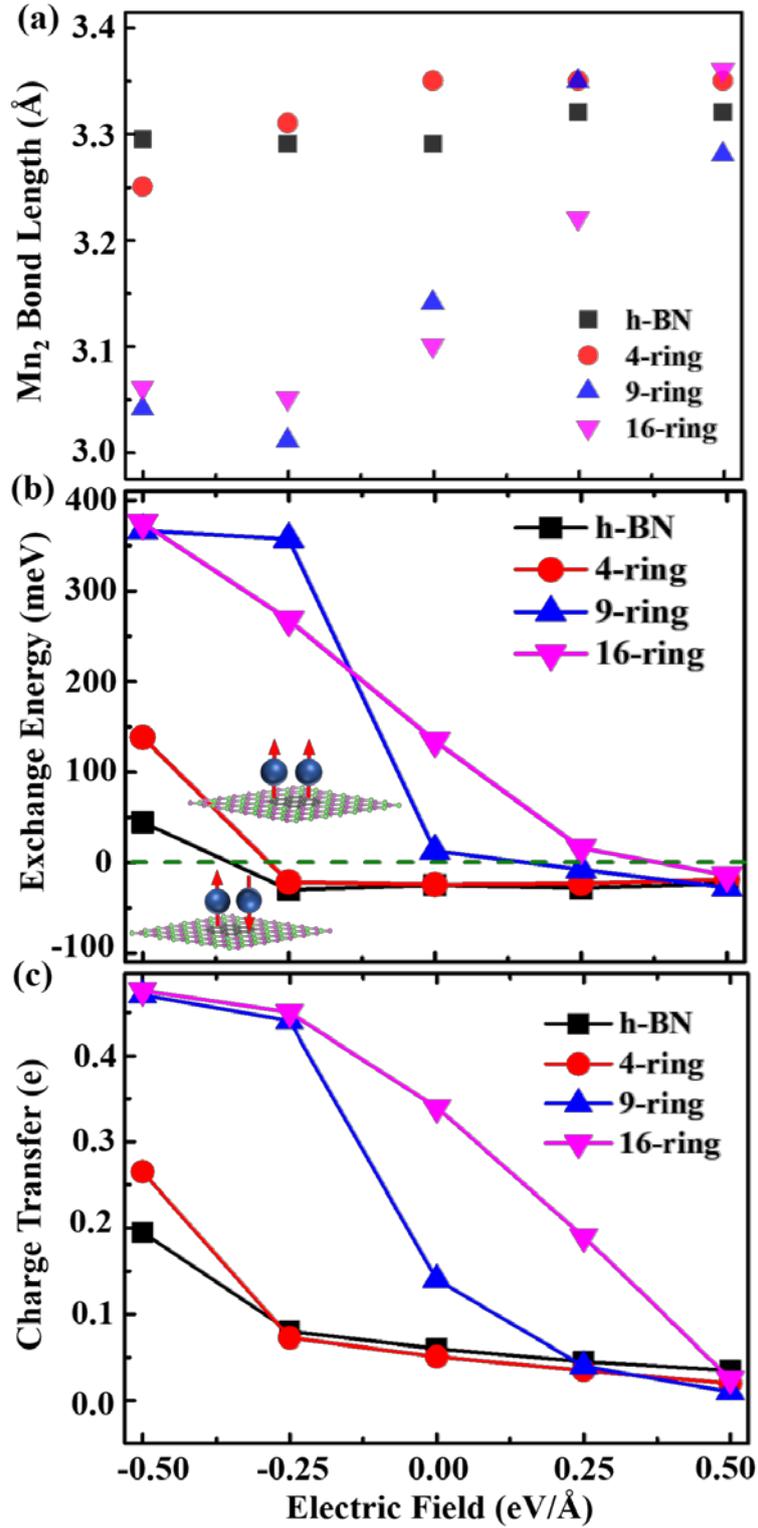

**Figure 4** Effect of electric field on (a) bond length Mn dimer, (b) exchange energy between FM and AFM states of Mn dimer, and (c) Charge transfer from Mn to the substrates for pristine h-BN and C-doped h-BN surfaces.

We now move on to discuss possible double exchange mechanism that arises due to the charge transfer to explain the FM ground state of Mn dimer on 16-C-ring-BN. For this we plot the density of states for Mn$_2$Cl and Mn dimer on 16-C-ring-BN in Fig. 5 (a) and (b). In Mn$_2$Cl one would naively expect three Cl *3p* orbitals for each of the up and down channel. In addition our calculation shows that there is prominent *pd* hybridization (between 3p electron of Cl atoms and 3d electrons of Mn atoms) in the up spin channel. Interestingly Cl 3p electrons are also hybridized with Mn 4s electrons. This implies that the Cl 3p electrons mediate the *sd* coupling (between Mn 3d up spin and 4s up spin electrons). For convenience we depict the above couplings using a schematic picture in the inset of Fig. 5(a) to explain the double exchange and the super exchange mechanisms. In this FM configuration 4s up electrons in both Mn atoms are immobile due to the Pauli's exclusion principle. But 4s down electron is free to move among the Mn atoms to gain the kinetic energy. In AFM configuration Mn moments interact antiferromagnetically to gain the super exchange energy but loses out on the kinetic energy. If the gain in kinetic energy, due to the indirect double exchange interaction, exceeds the super-exchange energy one expects a FM ground state. In fact FM is the ground state in Mn$_2$Cl (see Fig. 1). Similarly double exchange scenario also prevails between Mn atoms in Mn dimer on 16-C-ring-BN that helps the Mn moments to align ferromagntically.

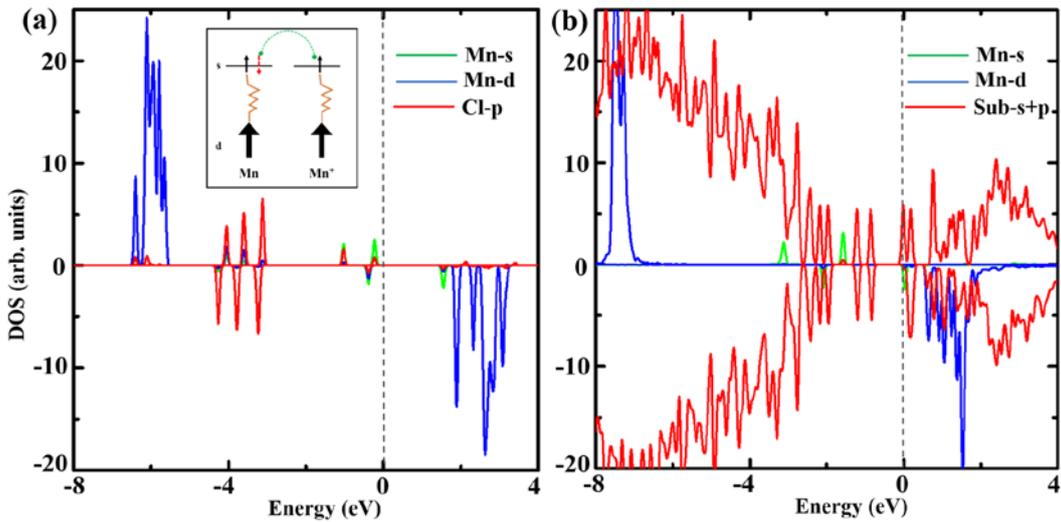

**Figure 5** Projected density of states of (a) Mn$_2$Cl and (b) Mn$_2$ on 16C-ring-h-BN surface. The Fermi level is set at 0 eV. Schematic diagram [inset of (a)] depicts the double exchange mechanism.

To get more insight about the charge redistribution at the interface of the dimer and the surface, it is worthwhile to investigate the charge density difference induced by the different substrates and can be expressed as follows: $\Delta n = n(S/Mn_2) - n(S) - n(Mn_2)$, where n(S/Mn$_2$), n(S), and n(Mn$_2$) are the total charge densities of the system consisting of Mn dimer placed above the 2D surface, corresponding 2D monolayer, and Mn dimer respectively. As shown in the Fig. 6(a), for the case of pristine h-BN surface, the large charge accumulation is distributed above the Mn atoms towards the vacuum and corresponding charge depletion region is distributed below and above the N atoms of the surfaces which lie in the planar region below the dimer. Due to small charge transfer from Mn atoms, a very small charge depletion region is seen just below the dimer and a respective charge accumulation region is situated in the space between the dimer and the 2D surface. Similar type of charge distribution is observed more prominently for 4C-ring-h-BN substrate [see Fig. 6(b)]. At higher concentration of C-rings in h-BN monolayer, the charge accumulation region surrounding the dimer starts decreasing indicating considerable amount of charge transfer from Mn atoms [see Figs. 6(c) and 6(d)] compared to the previous cases. For 16C-ring-h-BN substrate, there is significant charge redistribution in carbon rings just below the dimer by forming electron-affluent (in Mn dimer) and hole-affluent (in substrate) regions. This induces an intrinsic electric field, which directed from the substrate to the dimer resulting a considerable change in electronic and magnetic properties of the planar heterostructures.

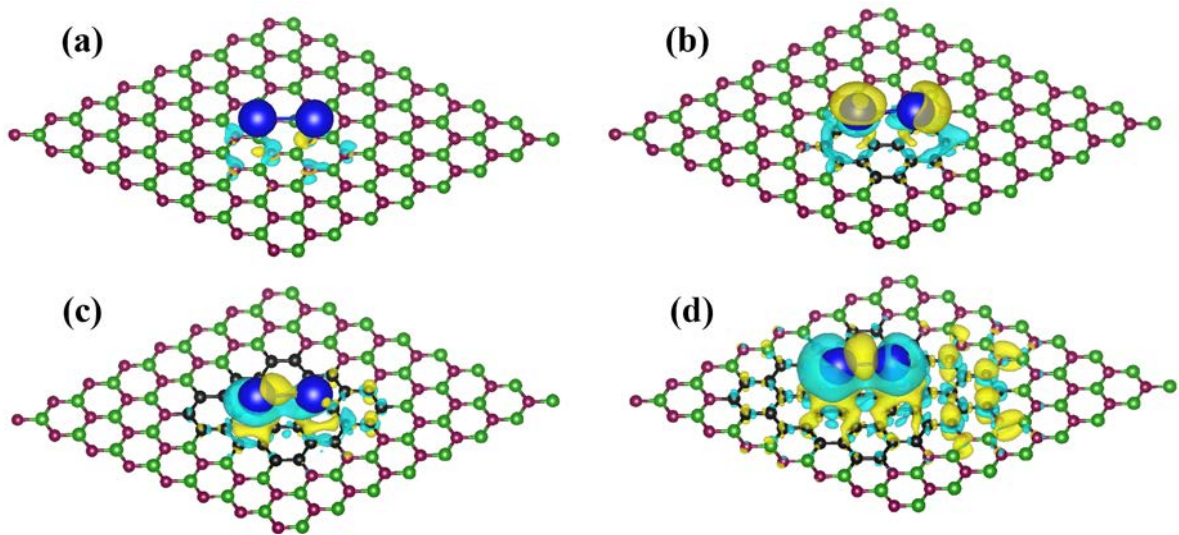

**Figure 6** Three-dimensional charge density difference plot of $Mn_2$ dimer on (a) h-BN, (b) 4C-ring-h-BN, (c) 9C-ring-h-BN, and (d) 16C-ring-h-BN surfaces. The value of iso-surface was taken as 0.0004 e/Å$^3$. Yellow and cyan regions represent electron accumulation and depletion regions respectively.

The effect of this internal electric field in $Mn_2$ on 16C-ring-h-BN surface can be enhanced or reduced by applying an external electric field (EEF). A positive electric field (applied from bottom to top direction) will negate the internal electric field and as a result the exchange energy will decrease. In fact, the exchange energy decreases with positive EEF [as shown in Fig. 4(b)] and the system switches to AFM for EEF=0.5 V/Å. The amount of charge transfer (from $Mn_2$ to the 2D surface) decreases [see Fig. 4(c)] with positive electric field, which supports the switching from the FM configuration to the AFM configuration. On the other hand, negative EEF (applied from top to bottom direction) enhances the charge transfer and increases the exchange energy. Bond length increases to 3.36 Å (for EEF=0.5 V/Å) from 3.12 Å (for EEF=0) while it decreases to 3.06 Å for EEF=-0.5 V/Å [see Fig. 4(a)]. Bond length (3.36 Å) obtained for EEF=0.5 V/Å is more or less equal to the $Mn_2$ bond length we get for the AFM configurations [see Fig. 1]. We obtained similar trend for $Mn_2$ on 9C-ring-h-BN surfaces as shown in Fig. 4. So, the charge redistribution due to application of EEF modifies the charge transfer between the Mn dimer and the 2D surface and as a result controls the magnetism.

As we discussed above Mn$_2$ prefers antiferromagnetic configurations in absence any external field in h-BN and 4C-ring-h-BN and the charge transfer (from Mn$_2$ to the 2D surface) is very small. So in positive electric field the charge transfer remains very small and the system remains antiferromagnetic [see Fig 4(b)]. In negative electric field, the charge transfer increases from Mn$_2$ to the 2D surface and systems prefer ferromagnetic configurations for EEF=0.5 V/Å. Corresponding charge transfer and Mn$_2$ bond length also supports this result (see Figs. 4(a) and (c)). So these results show that the application of electric field switches an AFM system to a FM system.

## Conclusion:

A number of perspectives with fundamental and application interests are opened by diversity of results presented in this work. With first principles density functional calculation, we have designed non-metallic 2D planar heterostructures as suitable substrates for Mn dimer for spin switch application. First we showed that the magnetic states of the dimer could be tuned by changing the doping concentrations of carbon in h-BN monolayer. Secondly we show that external electric field applied normal to the substrate can tune the magnetism. In our detailed calculations we unveil that the charge transfer controls the magnetism of Mn$_2$ dimer. It would be worthwhile for the experimenters to design small Mn cluster on carbon-doped-h-BN to tune the magnetism through electric field to device new nanoscale spintronics devices.

## Acknowledgement:

MRS and SKN would like to thank Centre of Excellence for Novel Energy Material (CENEMA) under the Ministry of Human Resources Development of India and School of Basic sciences, Indian institute of technology, Bhubaneswar, India.